\documentclass[10pt]{article}
\usepackage[OE]{express}
\usepackage{amsmath}
\usepackage{systeme}
\DeclareMathOperator{\sech}{sech}
\begin{document}

\title{Dispersion map induced energy transfer between solitons in optical fibres}

\author{A.\ Savojardo,\authormark{1,*} M.\ Eberhard,\authormark{2} and R.\ A.\ R\"{o}mer\authormark{1}}

\address{
\authormark{1}Department of Physics and Centre for Scientific Computing, The University
of Warwick, Coventry CV4 7AL, United Kingdom.\\
\authormark{2}Electrical, Electronic and Power Engineering, School of Engineering
and Applied Science, Aston University, Aston Triangle, Birmingham
B4 7ET, United Kingdom.\\
}

\email{
\authormark{*}a.savojardo@warwick.ac.uk\\
} 

\begin{abstract}
We propose an experimental setup to study collision-induced soliton amplification. In an optical fibre with anomalous dispersion ($\beta_2 < 0$), we replace a small region of the fibre with a normal dispersion fibre ($\beta_2 > 0$). We show that solitons colliding in this region are able to exchange energy. Depending on the relative phase of the soliton pair, we find that the energy transfer can lead to an energy gain in excess of $20\%$ for each collision. A sequence of such events can be used to enhance the energy gain even further, allowing the possibility of considerable soliton amplification. This energy transfer does not require a third order dispersion or Raman term.
\end{abstract}




\section{Introduction}

Solitons are spatially localized solutions that exist in many nonlinear wave equations \cite{Rem99}. Their collisions are fully elastic with all properties of both solitons conserved --- in particular their individual energies.
In optical fibres, soliton propagation is usually described using the non-linear Schr\"{o}dinger equation (NLSE) \cite{Rem99,Agr13}. The stability of the soliton solutions is the result of a perfect balance of dispersion and non-linearity \cite{ZakS72}. 
However, if this perfect balance is disturbed, more general collision scenarios become possible. Indeed, \emph{energy transfer} in pairwise soliton collisions has been observed experimentally \cite{MusKKL09,ArmCB15,BreSB16,DemAB14,GenSBD10} and discussed theoretically \cite{AkhK95,MusKKL09,ArmCB15,BreSB16,DemAB14,GenSBD10,EbeSMR2017}. 
This scenario usually requires the additional presence of third-order dispersion \cite{AkhSA10,EbeSMR2017,MiyN72}, Raman scattering \cite{LuaSYK06,DemAB14,ArmCB15,BreSB16}, and shock terms \cite{LuaSYK06,DemAB14,ArmCB15,BreSB16, Vei88} in the NLSE.

\begin{figure}[tb]
\begin{center}
\includegraphics[width=0.5\textwidth]{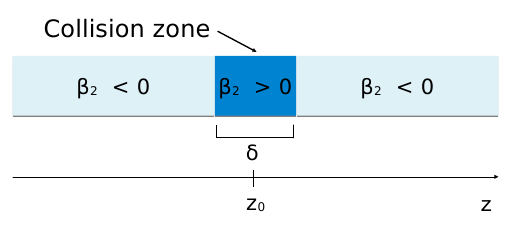}
\end{center}
\caption{\label{fig:Experimental-setup} Schematic representation of the optical fiber used for the thought experiment.  The horizontal fiber is indicated in blue. The light blue region corresponds to the anomalous-dispersion fiber ($\beta_2<0$) while the normal-dispersion fiber ($\beta_2>0$) of length $\delta$ is indicated in dark blue. The light propagation direction is indicated by a thin horizontal line including the point of collision $z_0$.}
\end{figure}

In this paper, we describe a dispersion map that leads to inelastic soliton collisions without any such additions. Consider a fibre with standard \emph{anomalous} dispersion in which two solitons, with different group velocities, propagate as stable pulses and eventually collide. 
The trick is now to replace the section of the fibre, where the soliton collision takes place, with a section of \emph{normal} dispersion fibre (cf.\ Fig.\ \ref{fig:Experimental-setup}). In the normal dispersion regime, the solitons are unstable, hence able to exchange energy under collision. Keeping this normal dispersion section short avoids excessive pulse-width broadening which would otherwise destroy the solitons. 
We optimize the dispersion map to maximize energy transfer between the solitons while keeping the disturbance of the soliton shape as small as possible. Hence two stable solitons emerge into the anomalous dispersion regime and continue to propagate.

\section{The model}

To model the situation proposed above, we use the NLSE \cite{Agr13} with a dispersion map $\beta_{2}(z)$
\begin{equation}
\partial_{z}u+\frac{i\beta_{2}(z)}{2}\partial_{t}^{2}u-i\gamma| u|^{2}u=0.\label{eq:NLSE}
\end{equation}
Here $u(z,t)$ is the pulse envelope, $z$ is the distance of propagation, $t$ is the time in the frame moving with the average group velocity of the carrier wave and $\gamma$ the non-linear coefficient. The dispersion $\beta_{2}(z)$ is equal to a constant $\beta_2>0$ in the region $\left[z_{0}-\frac{\delta}{2},z_{0}+\frac{\delta}{2}\right]$ and equal to $-\beta_2$ elsewhere; $\delta$ denotes the length of the normal ($\beta_2>0$) fibre symmetrically around the point of collision $z_{0}$. We start with an initial condition consisting of two solitons at $z\ll z_{0}$, such that we can write initially 
\begin{equation}
u\left(z,t\right)=\sqrt{P}_{1}\sech\left(\frac{t-t_{1}}{T_{1}}\right)e^{-i\varOmega_{1}t}+\sqrt{P}_{2}\sech\left(\frac{t-t_{2}}{T_{2}}\right)e^{-i\varOmega_{2}t+i\phi}.
\end{equation}
The two pulses have a relative phase difference $\phi$, and are characterized by power $P_{i}$, time $t_{i}$ and frequency shift $\varOmega_{i}$ for each pulse $i=1,2$. From these initial conditions we can calculate their inverse velocity $v_{i}^{-1}=\beta_{2}\Omega_{i}$, period $T_{i}=\sqrt{\frac{|\beta_{2}|}{\gamma P_{i}}}$ and energy $E_{i}=2P_{i}T_{i}$.

\section{Numerical results}

\begin{figure}[tb]
\begin{center}
{\scriptsize{}(a)}\includegraphics[width=0.47\textwidth]{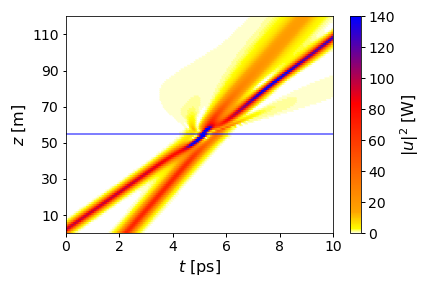}
{\scriptsize{}(b)}\includegraphics[width=0.47\textwidth]{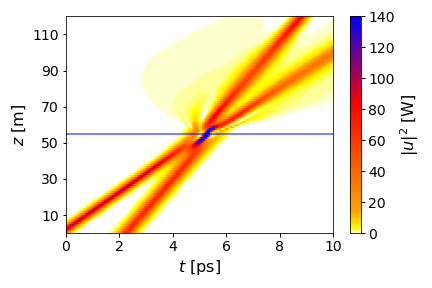}
\end{center}
\caption{\label{fig:Two-soliton-collision}Two soliton collision for the experimental setup described in Fig.\ \ref{fig:Experimental-setup}. The intensity $|u|^{2}$ is plotted as function of time $t$ and distance $z$. The phase difference $\phi$ corresponds to (a) the maximum and (b) minimum energy transfer $\Delta E_{1}/E_{2}$ plotted in Fig.\ \ref{fig:Energy-transfer-delta} (a). The horizontal blue line indicates the point of collision $z_0$ in the normal-dispersion fiber.}
\end{figure}

\begin{figure}[tb]
\begin{center}
(a)\includegraphics[width=0.45\textwidth]{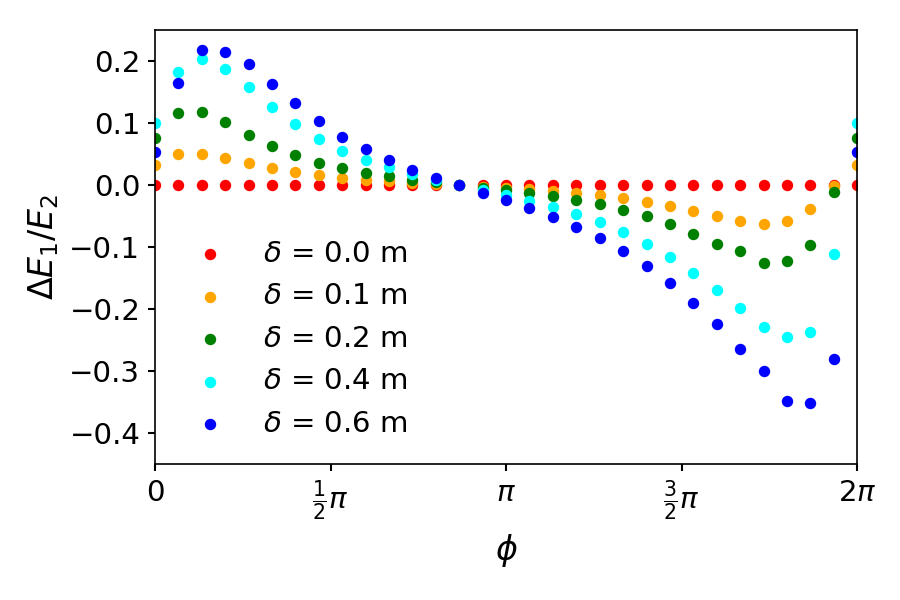}
(b)\includegraphics[width=0.45\textwidth]{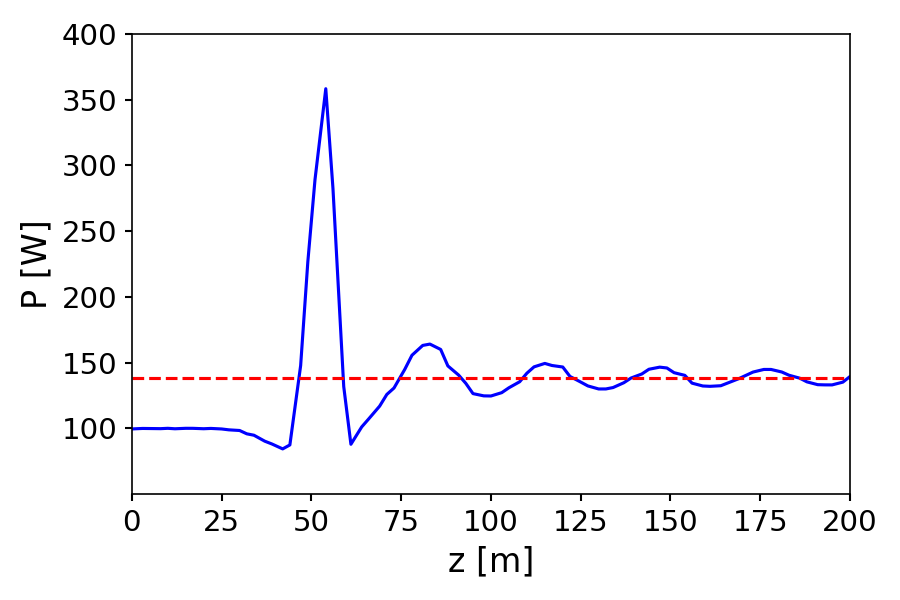}
\end{center}
\caption{\label{fig:Energy-transfer-delta}(a) Energy transfer $\Delta E_{1}/E_{2}$ as function of the phase difference $\phi$ for various lengths $\delta$ of the normal-dispersion fiber. (b) Maximum peak power (continuous blue line) as function of the distance $z$, calculated for the intensities in Fig.\ \ref{fig:Two-soliton-collision} (a). The dotted red line indicates the average power of $138$W after the collision.  }
\end{figure}

Fig.\ \ref{fig:Two-soliton-collision} (a) and \ref{fig:Two-soliton-collision} (b) show representative results for the collision of two solitons for the experimental setup described in Fig.\ \ref{fig:Experimental-setup}. The intensity $| u|^{2}$ is plotted as function of time $t$ and distance $z$. The initial conditions in both collisions are identical apart from the relative phase $\phi$ between the two solitons, in (a) $\phi=0.13\pi$ and in (b) $\phi=1.87\pi$. The initial power and frequency are $P_{1}=100$W, $\varOmega_{1}=-10$THz, $P_{2}=70$W and $\varOmega_{2}=-6$THz. The length $\delta$ of the optical fiber with normal dispersion is 0.6m. The optical fiber specifications are $|\beta_{2}|=0.01$ps$^{2}$m$^{-1}$ and $\gamma=0.003$W$^{-1}$m$^{-1}$.
%
In Fig.\ \ref{fig:Two-soliton-collision}, a first soliton (100W) collides with a second soliton (70W). After the collision both solitons emerge with peak power different from the initial one. In Fig. \ref{fig:Two-soliton-collision} (a) energy is transfered from the second to the first soliton, the emerging pulses have peak power of $\sim$ 138W and $\sim$ 24W respectively, the first soliton has gained 38W in amplitude. In Fig.\ \ref{fig:Two-soliton-collision} (b) we observe the opposite behavior, the emerging pulses have peak power of $\sim$ 50W and $\sim$ 73W respectively, although in this case the second soliton gains just 3W and half of the power of the first soliton is radiated.
%
{In Fig.\ \ref{fig:Energy-transfer-delta} (b) we show the maximum peak power of Fig.\ \ref{fig:Two-soliton-collision} (a) in order to highlight details of the process. The initial peak value is $100$W, then after the collision it oscillates and slowly dampens to the value $138$W when away from the collision region. This peak power mostly corresponds to the value of the first soliton.}

The numerical experiment confirms our hypothesis, instability leads to soliton energy exchange. Surprisingly energy transfer occurs without third order dispersion and Raman term \cite{LuaSYK06,AkhSA10}. The artificial instability, due to the change of sign in the second order dispersion term, is sufficient to mimic the effect of higher order terms for the two soliton collision.
To determine the influence of the phase difference $\phi$ and the length $\delta$ on the energy transfer, we studied collisions varying these two parameters.
For each value of $\phi$ and $\delta$, we calculate the percentage of energy transfer $\Delta E_{1}/E_{2}$, from soliton 2 to soliton 1. Fig.\ \ref{fig:Energy-transfer-delta} (a) shows the result of these calculations.
The energy transfer changes with $\phi$, it can be positive or negative meaning that energy can go from the second soliton to first one and vice versa. The local maximum and minimum are at $\phi=0.13\pi$ and $\phi=1.87\pi$ respectively. The maximum value of $\Delta E_{1}/E_{2}$ corresponding to the collision in Fig.\ \ref{fig:Two-soliton-collision} (a) is $\sim+22\%$, the minimum corresponding to the collision in Fig.\ \ref{fig:Two-soliton-collision}  (b) is $\sim-35\%$. This asymmetry towards the minimum is an indication of energy loss and can be understood as energy that is radiate during the scattering process. 
The energy transfer increases with $\delta$ and it is null when $\delta$ is zero, confirming that the unstable region is fundamental for the process to occur.

\section{Optical amplification }

\begin{figure}[tb]
\begin{center}
\includegraphics[width=0.6\textwidth]{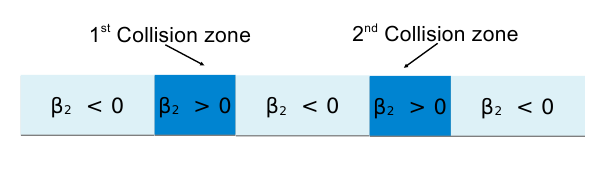}
\end{center}
\caption{\label{fig:Device-x2}Experimental setup for two consecutive amplifications. One normal-dispersion segment is placed at every point of collision. The colors are chosen as in Fig.\ \ref{fig:Experimental-setup}.}
\end{figure}

\begin{figure}[tb]
\centering{}
(a)\includegraphics[width=0.45\textwidth]{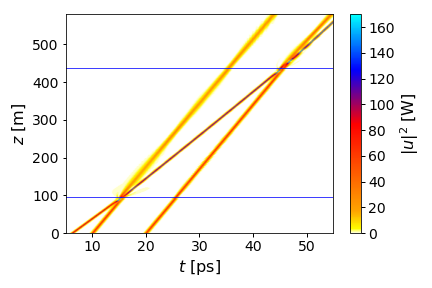}
(b)\includegraphics[width=0.45\textwidth]{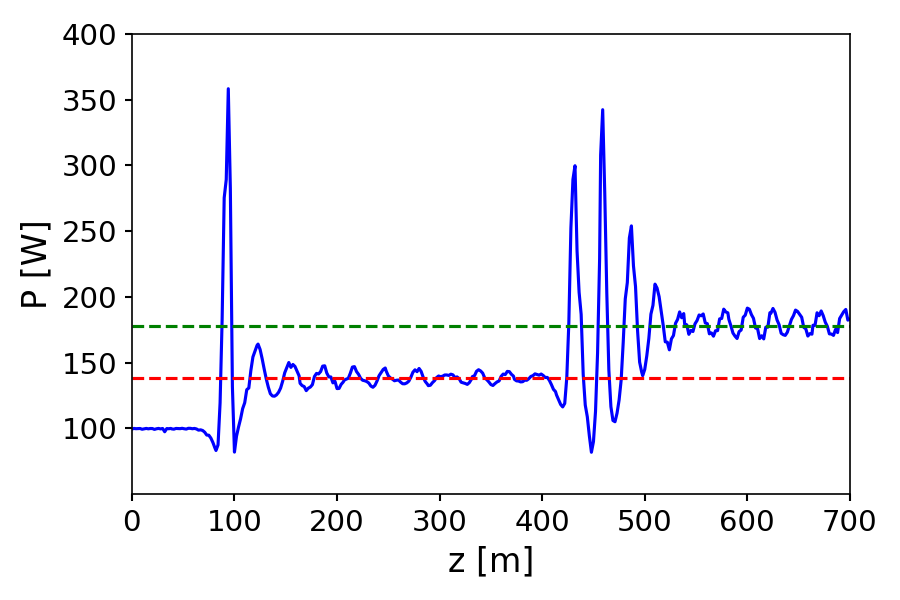}\\[-23ex]
\vspace*{-0.22cm}
\hspace*{6.25cm}
\includegraphics[width=0.14\textwidth,clip]{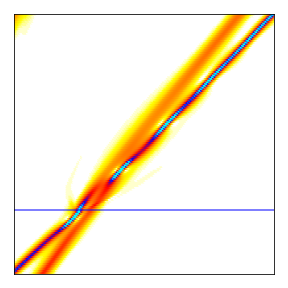}\\[10ex]
\caption{\label{fig:Two-amplifications} (a) Intensity $|u|^{2}$ plotted as function of time $t$ and distance $z$ for a three soliton collision process. An early soliton (starting at $z=0$ m and $t=4$ ps) is amplified twice in consecutive collisions. The horizontal blue lines indicate the points of collision in the normal-dispersion fiber. (b) Maximum peak power (continuous blue line) as function of the distance $z$, calculated for the collisions in (a). The dotted red and green lines indicate the average power of $138$W and $178$W after the first and second collision respectively. The inset shows details of the second collision for times $42$ps $\leq t \geq$ $54$ps and distances $400$m $\leq z \geq$ $550$m.}
\end{figure}

The above results show that the system in Fig.\ \ref{fig:Experimental-setup} is an amplification device. Particularly, assembling many of such devices in series a soliton can be amplified several times. For example, with three initial solitons and two devices in series like in Fig.\ \ref{fig:Device-x2} , we get the amplification shown in Fig.\ \ref{fig:Two-amplifications} (a). 
%
%
In this case a first soliton (100W) collides with a second (70W) and then a third soliton (70W) absorbing energy at every collision. 
{The first soliton emerges with an amplitude of $\sim$138W from the first collision and an amplitude of $\sim$178W from the second collision as shown in Fig. \ref{fig:Two-amplifications} (b)}. The initial phase differences are chosen so that the energy transfer is maximized at every scattering. In principle, there is no limit to the number of amplification devices that can be assembled and the early soliton can be amplified at every collision as long as it has a certain phase difference with the other colliding pulses.

\section{Conclusions}

Our numerical experiment indicates  the energy transfer between two unstable bright solitons in the normal-dispersion regions of a fiber. We find that the width $\delta$ of the normal dispersion region is crucial for the process to occur, as well as the phase difference $\phi$ between the two solitons. The device in Fig.\ \ref{fig:Experimental-setup} can be built in a real world laboratory and our predictions for the energy transfer (Fig.\ \ref{fig:Energy-transfer-delta} (a)) can be verified experimentally. Not only do our findings have fundamental implications in the field of soliton interaction and rogue waves generation \cite{AkhSA10,SolRKJ07} but the device in Fig.\ \ref{fig:Experimental-setup} can be exploited for soliton amplification in optical fiber systems.


\appendix

\section{Fit for the energy transfer function}

\begin{figure}[tb]
\begin{centering}
\includegraphics[scale=0.3]{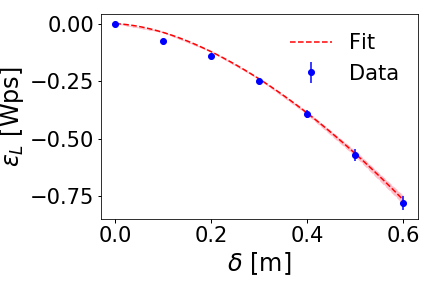}\includegraphics[scale=0.3]{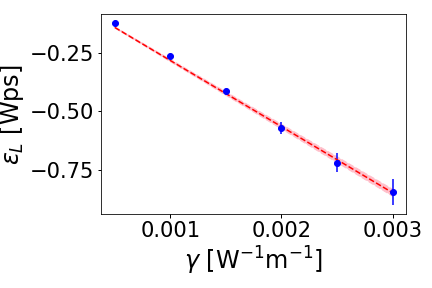}\includegraphics[scale=0.3]{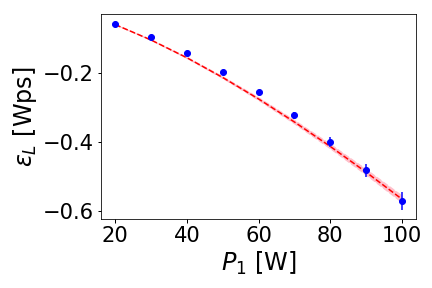}
\end{centering}
\begin{centering}
\includegraphics[scale=0.3]{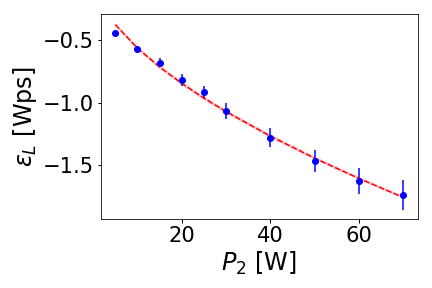}\includegraphics[scale=0.3]{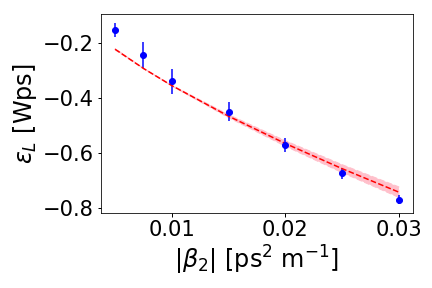}\includegraphics[scale=0.3]{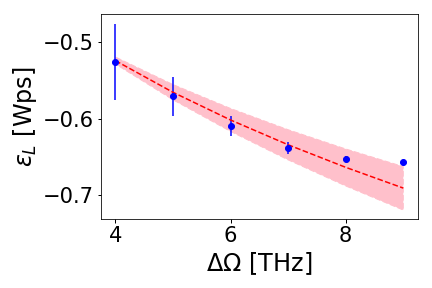}
\end{centering}
\caption{\label{fig:Fit-a-Plot}Functional dependence of $\epsilon_L$ on each of the six parameters given by Eq.\ (\ref{eq:set-parameters}). The data points represents numerical results of Eq.\ (\ref{eq:NLSE}) while the red lines denote the fit (\ref{eq:fit-a}), the pink region indicates the $68\%$ confidence level on the fit.}
\end{figure}

\begin{figure}[bt]
\begin{centering}
\includegraphics[scale=0.3]{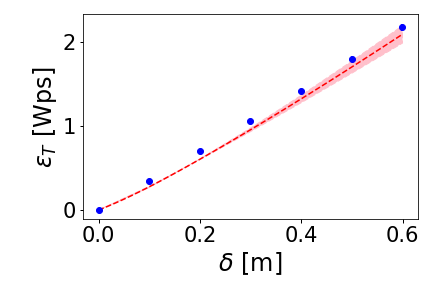}\includegraphics[scale=0.3]{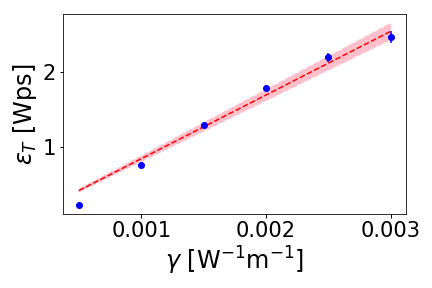}\includegraphics[scale=0.3]{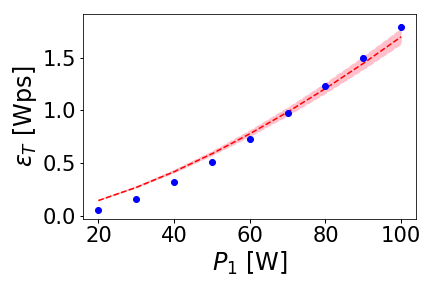}
\par\end{centering}

\begin{centering}
\includegraphics[scale=0.3]{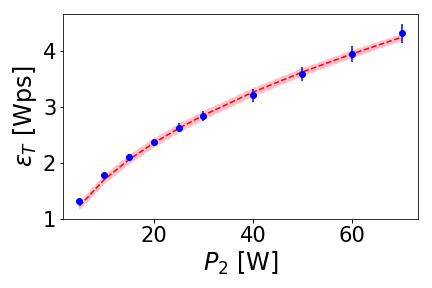}\includegraphics[scale=0.3]{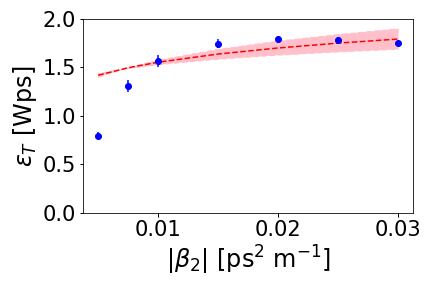}\includegraphics[scale=0.3]{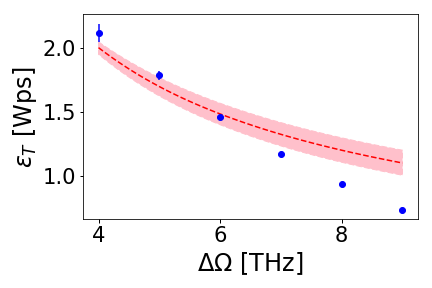}
\par\end{centering}

\caption{\label{fig:Fit-b-Plot}Functional dependence of $\epsilon_T$ on each of the six parameters given by Eq.\ (\ref{eq:set-parameters}). The data points represents numerical results of Eq.\ (\ref{eq:NLSE}) while the red lines denote the fit (\ref{eq:fit-b}), the pink region is as in Fig.\ \ref{fig:Fit-a-Plot}, indicating the $68\%$ confidence levels.}
\end{figure}

For the collisions in Fig.\ \ref{fig:Two-soliton-collision}  and \ref{fig:Energy-transfer-delta} we used two solitons with initial powers $P_1=100$W and $P_2=70$W. 
\begin{figure}[tb]
\begin{center}
\includegraphics[width=0.7\textwidth]{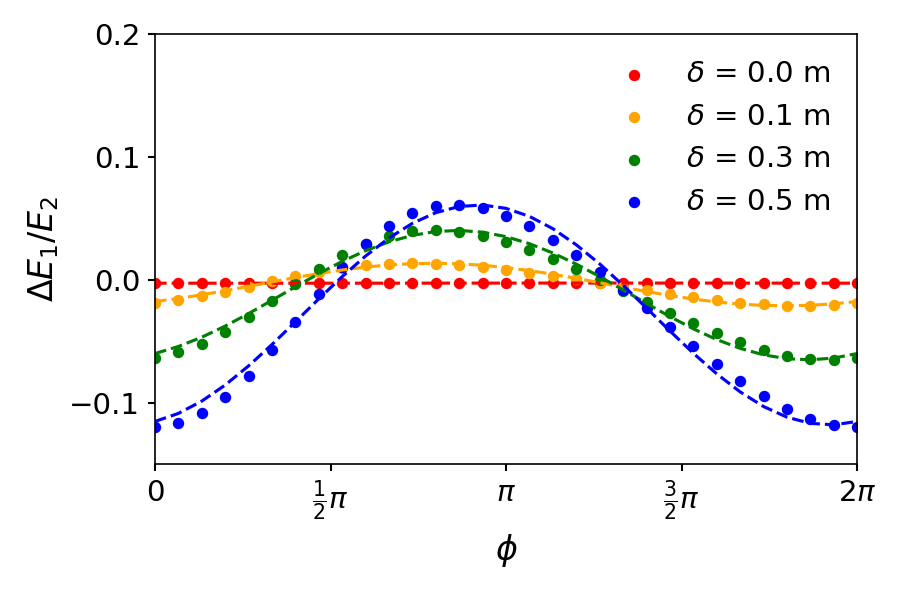}
\end{center}
\caption{\label{fig:Energy-transfer}Energy transfer $\Delta E_1/E_2$ as function of the phase difference $\phi$ and the length $\delta$. The data points represents result of the simulations while the lines denote the fit according to Eq.\ \eqref{eq:energy-fit}.}
\end{figure}
When $P_2 \ll P_1$, the energy transfer can be approximated as 
\begin{equation}
\triangle E_{1}=\epsilon_L+\epsilon_T\sin\left(\phi-\phi_{0}\right),\label{eq:energy-fit}
\end{equation}
where $\epsilon_L$, $\epsilon_T$ and $\phi_{0}$ are coefficients to be fitted.
Fig.\ \ref{fig:Energy-transfer}  shows a representative result for $P_1=100$W and $P_2=10$W. The data points represent results of the simulations while the lines denote a fit with \eqref{eq:energy-fit}.
The fit coefficients depend on the parameters 
\begin{equation}
\{\delta,\beta_{2},\gamma,P_{1},\varOmega_{1},P_{2},\varOmega_{2}\}.\label{eq:set-parameters}
\end{equation}
A dimensional analysis shows that $\epsilon_L$ and $\epsilon_T$ can be written in the form
\begin{subequations}
\begin{align}
\epsilon_L &=g_{1}\gamma\delta^{\lambda_{1}}|\beta_{2}|^{\lambda_{1}-1}\left(\varDelta\varOmega\right)^{2\lambda_{1}-3}P_{1}^{\eta_{1}}P_{2}^{2-\eta_{1}},\label{eq:fit-a}\\
\epsilon_T &=g_{2}\gamma\delta^{\lambda_{2}}|\beta_{2}|^{\lambda_{2}-1}\left(\varDelta\varOmega\right)^{2\lambda_{2}-3}P_{1}^{\eta_{2}}P_{2}^{2-\eta_{2}}.\label{eq:fit-b}
\end{align}
\end{subequations}
The coefficients $g_{1}$ and $g_{2}$ are dimensionless. Note that $\lambda_{1}$, $\eta_{1}$, $\lambda_{2}$ and $\eta_{2}$ are dimensionless by construction and $\Delta\varOmega=|\varOmega_{1}-\varOmega_{2}|$. In order to fit Eq.\ (\ref{eq:fit-a}) and (\ref{eq:fit-b}), we have calculated the energy transfer for a number of initial conditions by varying the set of parameters (\ref{eq:set-parameters}). We find the following values for the fit coefficients $g_{1}=-2.74\pm0.08$, $\lambda_{1}=1.67\pm0.02$, $\eta_{1}=1.417\pm0.007$, $g_{2}=3.01\pm0.19$, $\lambda_{2}=1.13\pm0.04$ and $\eta_{2}=1.529\pm0.015$. 
Fig. \ref{fig:Fit-a-Plot} and \ref{fig:Fit-b-Plot} show the fits of $\epsilon_L$ and $\epsilon_T$ using Eq.\ (\ref{eq:fit-a}) and (\ref{eq:fit-b}). The six dimensional fits are projected into two dimension at constant values $\delta=0.5$m, $|\beta_{2}|=0.02$ps$^{2}$m$^{-1}$, $\gamma=0.002$W$^{-1}$m$^{-1}$, $P_{1}=100$W, $\varOmega_{1}=-10$THz, $P_{2}=10$W and $\varOmega_{2}=-5$THz. 
The multidimensional fit and the numerical results are in an overall good agreement. 
{While a fit using rational exponents such as $\lambda_{1}=\lambda_{2}=\eta_{1}=\eta_{2}=1.5$ (or $\lambda_{1}=\eta_{1}=\eta_{2}=1.5$ and $\lambda_{2}=1$) is possible, the results are  much worse.}

\section*{Funding}
We are grateful to the EPSRC for provision of computing resources through the
MidPlus Regional HPC Centre (EP/K000128/1), and the national facilities
HECToR (e236, ge236) and ARCHER (e292).  We thank the Hartree Centre for use
of its facilities via BG/Q access projects HCBG055, HCBG092, HCBG109.

\section*{Acknowledgments}
We thank Akihiro Maruta for stimulating discussions in the early stages of the project. We are grateful to the  Centre for Scientific Computing (CSC), Warwick, for provision of computing resources.

\end{document}